\documentclass[conference]{IEEEtran}

\IEEEoverridecommandlockouts
\usepackage{cite}
\usepackage{amsmath,amssymb,amsfonts}
\usepackage{algorithmic}
\usepackage{graphicx}
\usepackage{textcomp}
\usepackage{xcolor}
\usepackage{subfigure}
\usepackage{caption}
\usepackage{url}
\usepackage{array,booktabs,arydshln,xcolor}
\usepackage{fancyhdr}
\usepackage{balance}

\def\BibTeX{{\rm B\kern-.05em{\sc i\kern-.025em b}\kern-.08em
    T\kern-.1667em\lower.7ex\hbox{E}\kern-.125emX}}

\makeatletter
\newcommand*\titleheader[1]{\gdef\@titleheader{#1}}
\AtBeginDocument{%
  \let\st@red@title\@title
  \def\@title{%
    \bgroup\normalfont\large\centering\@titleheader\par\egroup
    \vskip1.5em\st@red@title}
}
\makeatother

\newcommand{\graphwidth}{0.95\linewidth}
\newcommand{\vheightsave}{-2mm}

\title{Towards Traitor Tracing in Black-and-White-Box DNN Watermarking with Tardos-based Codes
\thanks{Work partially funded by the Spanish Ministry for Science and Innovation; Spanish Cybersecurity Institute (INCIBE) and NextGenerationEU/PRTR through projects "FELDSPAR: Federated Learning with Model Ownership Protection and Privacy Armoring" (Grant MCIN/AEI/10.13039/501100011033)  and "TRUFFLES: Trusted Framework for Federated Learning Systems", and by Xunta de Galicia and the European Regional Development Fund, under project ED431C 2021/47.}\vspace{-2mm}
}

\titleheader{\tiny © 2023 IEEE.  Personal use of this material is permitted.  Permission from IEEE must be obtained for all other uses, in any current or future media, including reprinting/republishing this material for advertising or promotional purposes, creating new collective works, for resale or redistribution to servers or lists, or reuse of any copyrighted component of this work in other works. \\ \vspace{-11mm}}

\author{\IEEEauthorblockN{Elena Rodríguez-Lois, Fernando Pérez-González}
\IEEEauthorblockA{
\textit{Signal Theory and Communications Department}\\
atlanTTic Research Center, University of Vigo, E.E. de Telecomunicación, 36310 Vigo, Spain\\
{erodriguez, fperez}@gts.uvigo.es\\\vspace{-8mm}}
}

\begin{document}

\maketitle

\begin{abstract}
The growing popularity of Deep Neural Networks, which often require computationally expensive training and access to a vast amount of data, calls for accurate authorship verification methods to deter unlawful dissemination of the models and identify the source of the leak. In DNN watermarking the owner may have access to the full network (white-box) or only be able to extract information from its output to queries (black-box), but a watermarked model may include both approaches in order to gather sufficient evidence to then gain access to the network. Although there has been limited research in white-box watermarking that considers traitor tracing, this problem is yet to be explored in the black-box scenario. In this paper, we propose a black-and-white-box watermarking method for DNN classifiers that opens the door to collusion-resistant traitor tracing in black-box, exploiting the properties of Tardos codes, and making it possible to identify the source of the leak before access to the model is granted. While experimental results show that the method can successfully identify traitors, even when further attacks have been performed, we also discuss its limitations and open problems for traitor tracing in black-box.
\end{abstract}


\section{Introduction and Previous Works} \label{sec:introduction}

With the recent advances in computational capacity and the great availability of training data, deep learning has revolutionized the way we approach many problems, both in industry and in our day to day life. But the training of Deep Neural Networks (DNNs) is usually time consuming, expensive, and may require private or proprietary data, making it unattainable to the general public. The development of pre-trained models, for users to fine-tune to their specific tasks, and, more interestingly, Machine Learning as a Service (MLaaS), has emerged in response to this demand as a way to capitalize DNNs. And thus, also born is the need to protect the Intellectual Property (IP) of models for deterring bad actors from unlawful use or dissemination. To this end, not only is it important to be able to prove the model's ownership, but also to determine the source of the leak, as it allows the owner to take further action. This source analysis becomes more challenging if bad actors are able to collude and produce models from multiple sources, which unfortunately is not difficult to imagine, if we consider that they would only need to pose as different clients to carry out this attack.

Forensic watermarking of DNNs has been presented as a promising solution to this problem \cite{Li21}, where a distinct watermark is inserted into the network at training time, so that it can be detected and used as proof of ownership later on. If one embeds unique watermarks, also known as fingerprints, into different copies, those can be used to identify the traitors that have unlawfully exploited or disseminated the model. The watermark itself can be embedded in the network parameters and intermediate features (white-box), meaning that the watermark detector must be granted access to the model, or in the input-output behaviour of the model (black-box), making the watermark detectable through queries. 

White-box schemes have greatly benefited from traditional media watermarking theory \cite{Li21,Barni21}. In one of the first white-box schemes for DNNs \cite{Uchida17}, the authors draw on the classic spread-spectrum and propose to embed an watermark into the model parameters through a regularization function during training, rewarding the model to converge towards a local minimum where the watermark can be correctly extracted from the model weights. Although other approaches have also been presented \cite{Li21}, this idea of regularization has been the most popular with many works building upon it, from the implementation of ST-DM watermarking \cite{Li21} to DNN fingerprinting with anti-collusion properties through Balanced Incomplete Block Design (BIBD) codes \cite{Chen19} \cite{Xu20}.

For black-box watermarking, previous works have mostly leveraged the same idea behind DNN backdooring attacks \cite{Barni21}. With this, during training a number of trigger-label pairs are injected into the model, to be eventually used to reveal the presence of the watermark, while the behaviour on other examples remains unaltered. The design of the trigger inputs has greatly varied across many works, using unrelated \cite{Li21}, benign \cite{Kallas22} or artificial \cite{Tekgul21} examples, adding visible and invisible patterns \cite{Li21}, and even using adversarial examples \cite{Li21}. While some existing works consider unique fingerprints to identify different model copies \cite{Fang-Qi_Li21,Maho23}, to the best of our knowledge, no scheme has been proposed for black-box watermarking that provides collusion-resistant traitor tracing capabilities. Thus, by keeping the model parameters secret from the legitimate author, colluders may easily succeed in using models unlawfully.  

White and black-box schemes are different in strengths and weaknesses. While white-box can be more reliable and hold much larger watermarking payloads, accessing the full network is unattainable in many cases. On the other hand, black-box allows for the detection of the watermark through querying the network, which may be available through MLaaS, but its capacity is restricted both by the output's dimensionality and the feasible number of queries: not only does the verification expose the triggers to the {\em spoil attack} \cite{Fang-Qi_Li21}, but it can also raise suspicion, and become expensive if the queries are available through a paid service. Because of this, these two ways can be seen as complementary \cite{Barni21}, and the use of both has already appeared in some previous works \cite{Darvish18,Bowen_Li22}.

Considering the above, the goal of this paper is two-fold: 
\begin{itemize}
    \item To show one can leverage previous watermarking theory, namely Tardos codes, to achieve traitor tracing capabilities in black-box, and that it is compatible with simultaneously fingerprinting in white-box. 
    \item To highlight the main challenges presented by black-box DNN fingerprinting that are yet to be solved, in order to motivate future research in this topic.
\end{itemize}

In what follows, Sec. \ref{sec:wm} describes the proposed scheme of both black and white-box approaches, Sec. \ref{sec:implement} explains the implementation of the experiments and presents the results, Sec. \ref{sec:discussion} includes a discussion about the most relevant open problems of black-box traitor tracing, and concludes this paper.  

{\em Notation:} Lower-case bold letters (e.g. $\textbf{a}$) represent column vectors, where their $i$th element is denoted with a subindex (e.g. $a_i$). Bold upper-case letters (e.g. $\textbf{B}$) represent two-dimensional matrices. Calligraphic fonts (e.g. $\mathcal{C}$) represent sets, alphabets or bases.

\section{Proposed Method} \label{sec:wm}

In the scenario considered by this work, a model owner trains a network on a classification task, which can later be disseminated among a number of known clients under a given user agreement. We assume the model owner is able to train different versions of the same network independently, so as to embed the unique watermarks that allow proof of ownership and source identification, even when users have colluded to produce a model from multiple sources. These should satisfy the requirements of \textit{Robustness}, \textit{Security}, \textit{Fidelity}, \textit{Capacity}, \textit{Integrity}, \textit{Generality} and \textit{Efficiency}, as defined in \cite{Li21}. 

To take advantage of both approaches, and ensure detectability without access to the model weights, the owner will embed a black-box and a white-box watermark simultaneously. With this, they can first gather sufficient proof of ownership through black-box queries to request access to the network, and later strengthen the accusation with a more thorough white-box watermark detection \cite{Barni21}. In case of a collusion of several users, one would want the black-box scheme to accuse at least one of the traitors (i.e. \textit{catch-one} goal in traitor tracing), and if successful, the white-box access can be used to find out about the remaining sources.

\subsection{Black-Box Fingerprinting with $q$-ary Tardos Codes} \label{subsec:black}

Similar to backdooring attacks, black-box DNN watermarking can embed information in the input-output behaviour, activated by certain queries performed by the model owner, who does not need access to the network parameters. As previously mentioned, one of its main disadvantages is the embedding capacity, which is limited by the dimensionality of the network's output. This is usually circumvented by employing a high number of triggers, which are memorized during the training of the model and ideally do not negatively impact the main task, so that the achieved outputs to the queries can be statistically significant while preserving the model's performance on normal users' queries. Still, utilizing as few triggers as possible to detect the watermark should be considered a design priority for black-box watermarking schemes: On the one hand, any trigger that is shown to a suspected model could be susceptible to a spoil attack \cite{Fang-Qi_Li21}, where the stolen model could now be trained to fit the known triggers on random labels, but also ignored, or even classified with a different non-copyrighted model, rendering them useless. On the other hand, queries may only be accessible through a paid service, making the accusation process expensive if many triggers are needed.

When it comes to black-box fingerprinting, if the same approach is to be followed, different copies of the model would need to be distinguishable through the output of the trigger queries. This was contemplated in a previous work for a federated learning watermarking framework \cite{Fang-Qi_Li21}, but unfortunately this approach does not consider any attacks specifically against the unique fingerprints (e.g., model averaging), which makes it straightforward for any bad actor to generate an untraceable copy of the model. While these attacks are easy to model in the white-box approach, the impact over the black-box triggers is difficult to predict. This makes strategy-dependent anti-collusion codes such as BIBD, previously used in white-box fingerprinting \cite{Chen19,Xu20}, less suitable. Fortunately, extensive research has been done on Tardos codes, which are collusion-resistant fingerprinting codes, agnostic to the attack strategy of the colluders. The original proposal only considered binary fingerprinting codes, but subsequent works have generalized this strategy to any $q$-ary alphabet \cite{Skoric12}, allowing us to leverage the full dimensionality of the network's output, and making these codes an interesting fit for this new fingerprinting scenario. Tardos codes mostly rely on the so-called Marking Assumption (MA), which states that attackers are not able to produce changes in sections of the content where they received the same information. In the current scenario, this would mean that if no colluder outputs a given symbol to a trigger, the merged model will not give that output either.\footnote{Otherwise, it known as the \textit{wide-case} problem \cite{Barg03} in traitor tracing.} Actually, the MA may not always hold for black-box triggers, due to unexpected effects of the merge on the model internal features or further attacks (see Sec. \ref{sec:discussion}), but fortunately, as results will show, one can still successfully utilize Tardos codes for this application. 


In traditional Tardos fingerprinting, the content would be divided into $m$ segments (or detectable positions), each hiding a symbol $\alpha$ of the message from a discrete alphabet $\mathcal{Q}$ of size $q$. For the sake of simplicity, we will keep the traditional notation for black-box DNN fingerprinting. The role of the different segments will be covered by the set of $m$ trigger examples, with each being assigned a label $\alpha$, from the possible outputs $\mathcal{Q}$. Each user $j$ will be assigned a unique fingerprint $\boldsymbol{x}_j$, where $x_{ji}$ specifies the label for trigger $i$.

As described in \cite{Skoric12}, $q$-ary Tardos codes are generated from a secret $q$-component bias vector $\boldsymbol{p}^{(i)}$ as $p_\alpha^{(i)} \doteq P[x_{ji}=\alpha]$. The bias vector $\boldsymbol{p}^{(i)}$ is independently drawn from a symmetric Dirichlet distribution:
\begin{equation} \label{eq:dirichlet}
    F(\boldsymbol{p}) = \frac{1}{\mathcal{N}(q,\kappa,\tau)}\prod_{\alpha\in\mathcal{Q}}p_\alpha^{-1+\kappa}
\end{equation}
where $\kappa>0$ is the concentration parameter, $\tau$ a cutoff parameter for $p_\alpha \in [\tau, 1-(q-1)\tau]$,\footnote{Although this is not necessary in the implementation of Tardos schemes for $q\ge 3$ the work in \cite{Skoric12} includes a cutoff parameter $\tau$ restricting $p_\alpha^{(i)} \in [\tau, 1-(q-1)\tau]$ that simplifies the theoretical analysis, which is also considered here so that equations derived from \cite{Skoric12} still apply.} and $\mathcal{N}(q,\kappa,\tau)$ is a normalization factor that guarantees that
\begin{equation} \label{eq:nqkt}
    \int_{\tau}^{1-(q-1)\tau} \cdots \int_{\tau}^{1-(q-1)\tau} \delta (1- \sum_{\alpha \in \mathcal{Q}} p_\alpha) F(\boldsymbol{p}) \text{d}\boldsymbol{p} = 1,
\end{equation}
with $\delta$ the Dirac Delta function.

In our work we will consider two different values of $\kappa$: a default $\kappa = \frac{1}{q}$, which is the recommended in \cite{Skoric12} to guarantee a positive guilty score regardless of the collusion strategy, and $\kappa\gg \frac{1}{q}$, which greatly favors the detection of strategies similar to majority-voting, but can potentially result in negative scores for guilty participants under minority-voting \cite{Simone11}. The value of $\tau$ is optimized for $\kappa = \frac{1}{q}$ as $\tau = c_0^{-2/(1+\kappa)}$ \cite{Skoric12}, where $c_0$ represents the maximum number of colluders considered by the scheme.


In traditional Tardos schemes an accusation score would be computed across all $m$ segments and compared to an accusation threshold, but as mentioned above, one should be careful not to expose too many triggers in black-box watermarking in order to mitigate spoil attacks \cite{Fang-Qi_Li21}. This means that ideally, for any suspected leak, the model owner should query with as few triggers as possible, $t^*$, that allows them to make a decision. After $t$ queries to the model, the author will have exposed a subset of $t$ trigger examples from the available $m$, and obtains an attack vector $\mathbf{y}=[y_1,...,y_t]^T$. For each suspected participant, the accusation score is computed in \cite{Skoric12} as $S_j=\sum_{i=1}^t S_j^{(i)}$, with

\begin{equation} \label{eq:segmentscore}
    S_j^{(i)}= 
        \begin{cases}
        U_1(p^{(i)}_{y_i}) & \text{if } x_{ji} = y_i \\
        U_0(p^{(i)}_{y_i}) & \text{if } x_{ji} \neq y_i
        \end{cases}   ,
\end{equation}
and where
\begin{align} \label{eq:scorefunctions}
    U_1(p)&=\sqrt{(1-p)/p},  & U_0(p)&=-\sqrt{p/(1-p)}.
\end{align}

Instead of the fixed accusation threshold used in the traditional schemes \cite{Skoric12}, with our proposed approach the owner can perform a Sequential Probability Ratio Test (SPRT) \cite{SPRT} with every trigger queried. This allows them to stop compromising triggers as soon as sufficient evidence is gathered, to either accuse or exonerate the suspected model. Since in our scenario the MA does not always hold (see Sec. \ref{sec:discussion}), the theoretical analysis for the scores of guilty participants in \cite{Skoric12} is no longer applicable. While adapting this analysis to encompass MA violations is left for future research, here we assume that the owner has an empirical estimate of the score distributions $S_j^{(i)}$. For this, we will define a collusion $\mathcal{C}^c$ of  $c \leq c_0$ users as the set of their indexes, and denote $\mathcal{Z}^c$ as the set of all possible collusions of size $c$. Then, the owner can randomly sample $\mathcal{Z}^{c_0}$ and obtain the statistical distributions of the scores when $j$ is a colluder and when it is not. Let 
$P_{\mathsf{col}}(S_j^{(i)})$ and $P_{\mathsf{inn}}(S_j^{(i)})$ respectively denote such distributions;
then, for each new observed score $S_{j}^{(t)}$, the cumulative sum of the log-likelihood ratio $W_j^{(t)}$ is computed as
\begin{equation} \label{eq:wald}
W_j^{(t)} = W_j^{(t-1)} + \text{log} \left(\frac{P_{\mathsf{col}}(S_{j}^{(t)})}{P_{\mathsf{inn}}(S_{j}^{(t)})}\right).
\end{equation}

For any user $j$, if $W_j^{(t)}$ exceeds a threshold $b$, then an accusation can be made. Alternatively, if $W_j^{(t)} < a$ $\forall j$ the model can be deemed innocent. In either case, a decision has been reached and $t^* = t$. These thresholds are set as $a\approx \text{log}(\epsilon_2/(1-\epsilon_1))$, $b\approx \text{log}((1-\epsilon_2)/\epsilon_1)$, where $\epsilon_1$ and $\epsilon_1$ represent the desired False Positive (FPR) and False Negative Rates (FNR), respectively. As previously mentioned, this would be analogous to the \textit{catch-one} goal in traitor tracing, where the author needs to identify at least one colluder, as this would constitute enough evidence to grant access to the network and perform a deeper white-box analysis.

An additional stop condition can be set as a function of $t$ and $\epsilon_1$, according to the bound on $\epsilon_1$ presented in \cite{Skoric12}:
\begin{align} \label{eq:pfpbound}
    \epsilon_1 \leq \text{exp}\left(\frac{Z_t^2}{2t} \cdot \frac{1}{1 + Z_t / (3t\sqrt{\tau})}\right),
\end{align}
where the author can also accuse a guilty user if $S_j>Z_t$.

\subsection{White-Box Fingerprinting with Orthogonal Codes} \label{subsec:white}

White-Box watermarking exploits the over-parametrization of DNNs to encode information into the internal values of the network, for example the weights, or activations maps of certain layers. While state-of-the-art architectures grow in parameters to better solve the main task, this also confers them with more capacity to host the watermark \cite{Barni21}, making white-box schemes ever more suitable for fingerprinting DNNs. The most popular approach, based on \cite{Uchida17}, relies on the use of a regularization function, leading the network to converge towards a local minimum of the main task where the watermarking task can also be satisfied. Previous works have already been presented \cite{Chen19,Xu20} that take this angle to successfully implement anti-collusion fingerprinting of DNNs with BIBD codes. In this paper we consider a similar idea, but that is not restricted to using binary vectors: embedding non-binary orthogonal codes, that form the basis of the vector space where traitors can be identified. This allows us to directly project the values found on the pirated model onto the vector basis, without binarizing them first, which would induce unintended noise. 


Following the approach in \cite{Uchida17}, in each of the copies of the model, a white-box fingerprint will be embedded into vector $\textbf{w}$ of the $l$ flattened parameters of a given layer, through a regularization function adding to the total loss $E(\textbf{w})$, as $E(\textbf{w}) = E_0(\textbf{w}) + \lambda E_R(\textbf{w})$, where $E_0(\textbf{w})$ represents the main task loss, $E_R(\textbf{w})$ the regularization loss, and $\lambda$ is a parameter controlling its strength during training. For this, the fingerprints are generated as a $p$-dimensional orthonormal basis $\mathcal{S} = \{\textbf{s}_1, \textbf{s}_2, ..., \textbf{s}_p\}$, where $\textbf{s}_j$ represents the $j$th user basis vector, and a secret matrix $\mathbf{D}$ of size $l \times p$ (sampled from a normal distribution), that projects the $l$ parameters of the layer onto the $p$ dimensions of the user basis. The goal of $E_R(\textbf{w})$ is to reward the model to update the layer weights towards a vector that maximizes the projection over the user basis vector, $\textbf{s}_j$, and it is computed as
\begin{align} \label{eq:regwb}
    E_R(\textbf{w}) = \exp \left(-\frac{\textbf{w}^\intercal \cdot \mathbf{D} \cdot \textbf{s}_j}{\|\textbf{w}^\intercal \cdot \mathbf{D}\|}\right).
\end{align}


Upon access to the weights of a suspected model, the projection on each user vector becomes
\begin{align} \label{eq:proj}
    r_j = \frac{\textbf{w}^\intercal \cdot \mathbf{D} \cdot \textbf{s}_j}{\|\textbf{w}^\intercal \cdot \mathbf{D}\|}.
\end{align}

The expectation on the projections if all colluders participate equally (and there are no further attacks) is 
\begin{align} \label{eq:eproj}
    \mathbb{E} \{r_{j|j\in \mathcal{C}^c}\} &= 1/\sqrt{c},  & \mathbb{E} \{r_{j|j\notin \mathcal{C}^c}\} &= 0,
\end{align}
allowing the author to effectively identify them.

\section{Implementation and Experimental Results} \label{sec:implement}

\subsection{DNN Architecture and Main Task}

In order to validate the traitor-tracing capabilities of the proposed scheme we have chosen an image classification task on the well-known MNIST dataset \cite{MNIST}. Neither the white or black-box approaches are architecture-dependent, so we have used a simple CNN for our experiments. However, bigger networks, with more parameters, would be expected to have even more capacity for both watermarks. The CNN is made up of three convolutional layers of 16, 64 and 128 3$\times$3 kernels, respectively, followed by a final fully-connected layer with 10 neurons with a softmax activation function, which maps to the 10 classes on MNIST. For all our experiments, we consider the output class to be the highest softmax neuron, regardless of the soft value of the output vector.

\subsection{Choice of trigger set} \label{sec:triggsetimp}

In previous black-box watermarking works, the importance of choosing a trigger set similar to that of the main task has been highlighted, as out-of-distribution triggers could be detected by the adversary when the model is queried \cite{Barni21}, deeming those triggers potentially useless. However, when models are merged by colluders, the output of triggers with different labels might be skewed towards the common main task, which could reduce the suitability of Tardos Codes. To assess this limitation, we have considered 3 different types of triggers:\looseness=-1

\begin{itemize}
    \item $\mathcal{T}_R^m$: Randomly generated patterns, from a uniform distribution.
    \item $\mathcal{T}_B^m$: Benign examples from the main task dataset MNIST, set apart and excluded from the main task training.
    \item $\mathcal{T}_M^m$: Merging of benign examples, not excluded from the main task training, with an invisible random pattern from a uniform distribution as $\mathcal{T}_{M_i} = 0.9 \cdot \mathcal{T}_{B_i} + 0.1 \cdot \mathcal{T}_{R_i}$.

\end{itemize}
The $\mathcal{T}_R^m$ trigger set will be used as baseline, and should be assumed if no trigger set is explicitly stated. For the chosen set, the labels for each user will be assigned according to $x_{ji}$, as described in Sec. \ref{subsec:black}.

\subsection{Watermarking Parameters and Training Process}

When training the different model copies, the networks learn not only the main task, but also the two watermarking tasks. Copies start from the same weight initialization,\footnote{It was found experimentally that different initializations do not allow traitor tracing for this black-box scheme, most likely because the overfitting on the same labels relies on features that are too different.} and are trained on the main task dataset for 10 epochs, with categorical cross-entropy loss, stochastic gradient descent with 0.001 learning rate, and a batch size of 16. From MNIST, 25\% of the examples are set apart to evaluate the model's performance on the main task, and also perform user attacks. Regarding the watermarking, the third convolutional layer was chosen to embed the white-box watermark, with 73,728 parameters. We chose a user basis of $p$=1000 dimensions, and the strength of the white-box regularization function $\lambda$ was set to 1. For the black-box watermark, the full trigger set is fitted with their respective labels every 100 batches of the main dataset, with the same learning parameters. The maximum number of triggers $m$ is 1000, $c_0$ was set to 6, and the resulting cutoff parameter $\tau$ is 0.038. The values of $\epsilon_1$ and $\epsilon_2$ were set to $10^{-6}$ and $10^{-3}$, respectively, leaving $a=-3$ and $b=6$ in the SPRT. All experimental values were obtained after training 100 model copies, and collusions of size $c$ were modeled by randomly sampling $\mathcal{Z}^c$ 500 times. The experimental distributions of the Tardos scores $P_{\mathsf{inn}}(S_j^{(i)})$ and $P_{\mathsf{col}}(S_j^{(i)})$ were calculated in previous simulations for each configuration, across all 1000 triggers.

\subsection{User Attacks on the Individual Watermarks}

This work mainly focuses on the collusion attack, where several users merge their individual models to create a new copy. For simplicity, we modeled the merging as an averaging of the model weights, but other collusion strategies could also produce similar results.\footnote{For example, randomly sampling the model parameters (\textit{interleaving} in traitor tracing) presented a similar behaviour in preliminary experiments.}  Users can further attack the merged copy by performing fine-tuning or pruning operations. In what follows, the fine-tuning attack fits the validation split for 10 epochs with the original training parameters, and the pruning attack uses the validation split to set 80\% of the kernels' weights to zero. Other attacks to the watermark are also possible, and should be studied in future work.

\subsection{Experimental Results} \label{subsec:expresult}

\subsubsection{Impact of Tardos Codes}

\begin{figure}[t]
    \centering
    \includegraphics[width=\graphwidth]{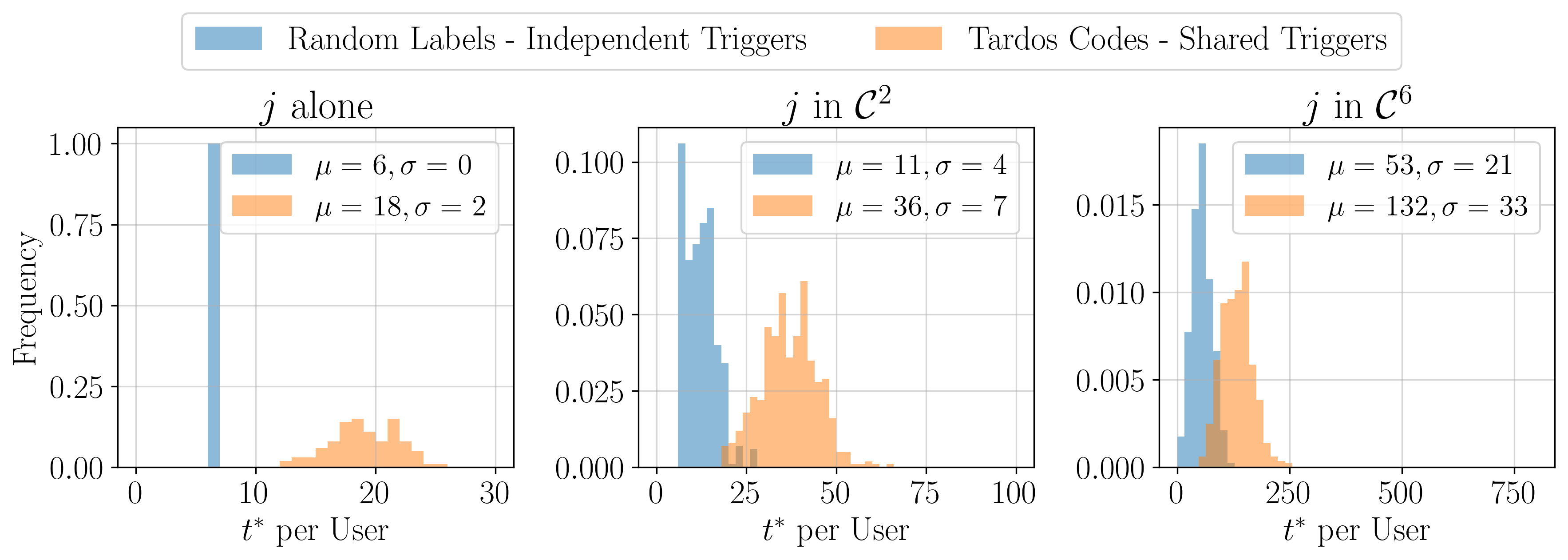}
    \caption{Impact of using Tardos codes on the number of queries needed for a single user before an accusation.}
    \label{fig:fanqli}
\vspace{\vheightsave}
\end{figure}

Fig. \ref{fig:fanqli} shows the histogram comparison of the number of queries $t^*$ needed per user, between the proposed scheme and the existing approach in \cite{Fang-Qi_Li21}, where independent triggers with random labels are assigned to each participant. In that case, querying stops as soon as the FPR, calculated as FPR$={t \choose t'}\cdot 0.1^{t'}\cdot (1-0.1)^{t-t'}$, with $0.1$ the probability of randomly answering a query correctly and $t'$ the number of correctly answered queries, falls below $\epsilon_1$, so the desired FPR is the same in both approaches. When analyzing Fig. \ref{fig:fanqli}, one must keep in mind that independent triggers will require queries for all possible participants, so the total number of queries will be $\#\text{users} \times t^*$, whereas for the shared triggers, the total number is still $t^*$. This makes the independent approach a more expensive, and potentially more suspicious accusation process. As an example, if 100 users have access to the model and one of them leaks it, one would need an average of 600 queries to identify the source with independent triggers, and only 18 with the proposed approach. Fortunately, the sharing of the triggers causes only a reasonable increase of $t^*$ per user for the smaller collusions, but still, the approach in \cite{Fang-Qi_Li21} would expose less triggers per user, making it slightly more suitable against the spoil attack.

\subsubsection{Influence of $\kappa$}
\begin{figure}[t]
    \centering
    \subfigure[$\kappa = 0.1$]{
        \label{fig:kappa0.1}
        \centering        \includegraphics[width=\graphwidth]{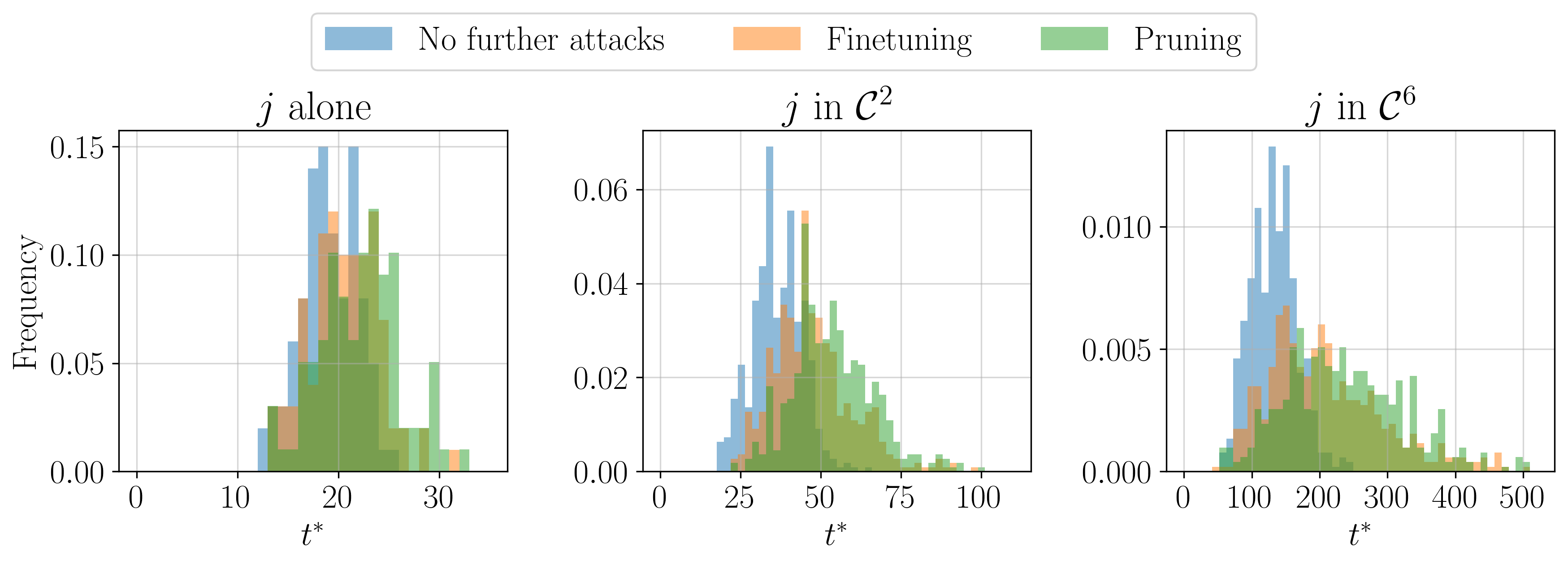}
    }
    \subfigure[$\kappa = 100$]{
        \label{fig:kappa100}
        \centering
        \includegraphics[width=\graphwidth]{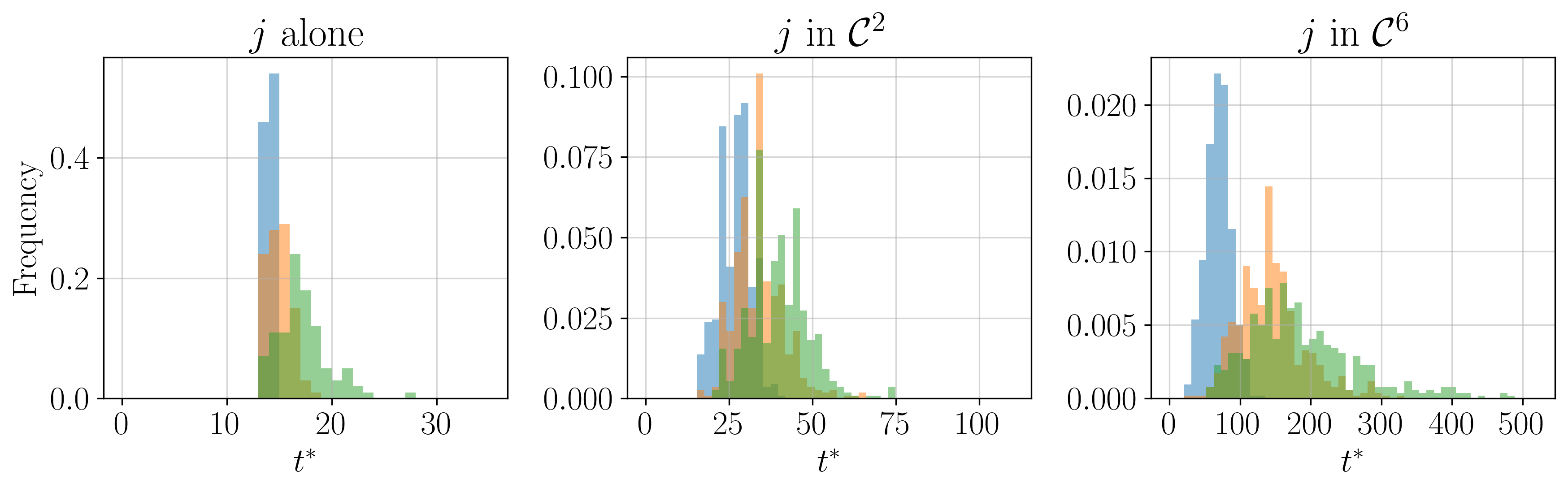}
    }
    \caption{Experimental distribution of $t^*$ according to $\kappa$. }
    \label{fig:kappat}
    \vspace{\vheightsave}
\end{figure}

As mentioned in Sec. \ref{subsec:black}, the recommended value of $\kappa = \frac{1}{q}$ \cite{Skoric12} guarantees that no collusion strategy results in a negative guilty Tardos score, but other values may be better suited to target specific colluding scenarios: Fig. \ref{fig:kappat} shows the impact of significantly increasing $\kappa$, which improves the efficiency of the scheme, even in the presence of further attacks, such as fine-tuning or pruning. The attacks, as expected, increase $t^*$, as the MA will be violated for a higher percentage of triggers (see Sec. \ref{sec:discussion}). Exonerating innocent models is also feasible under the SPRT, with an average of $t^*=301$ for $\kappa=0.1$, and $t^*=226$ for $\kappa=100$, implying again that a higher $\kappa$ may be better suited for this case. As mentioned in Sec. \ref{subsec:black}, the results for both values of $\kappa$ suggest that when averaging model weights, the resulting strategy over the trigger outputs may resemble majority-voting \cite{Simone11}. With that in mind, a deeper analysis, including other types of attacks, should help in the design of more targeted Tardos codes.\looseness=-1

\subsubsection{Influence of $\mathcal{T}$}
\begin{figure}[t]
    \centering
    \includegraphics[width=0.75\linewidth]{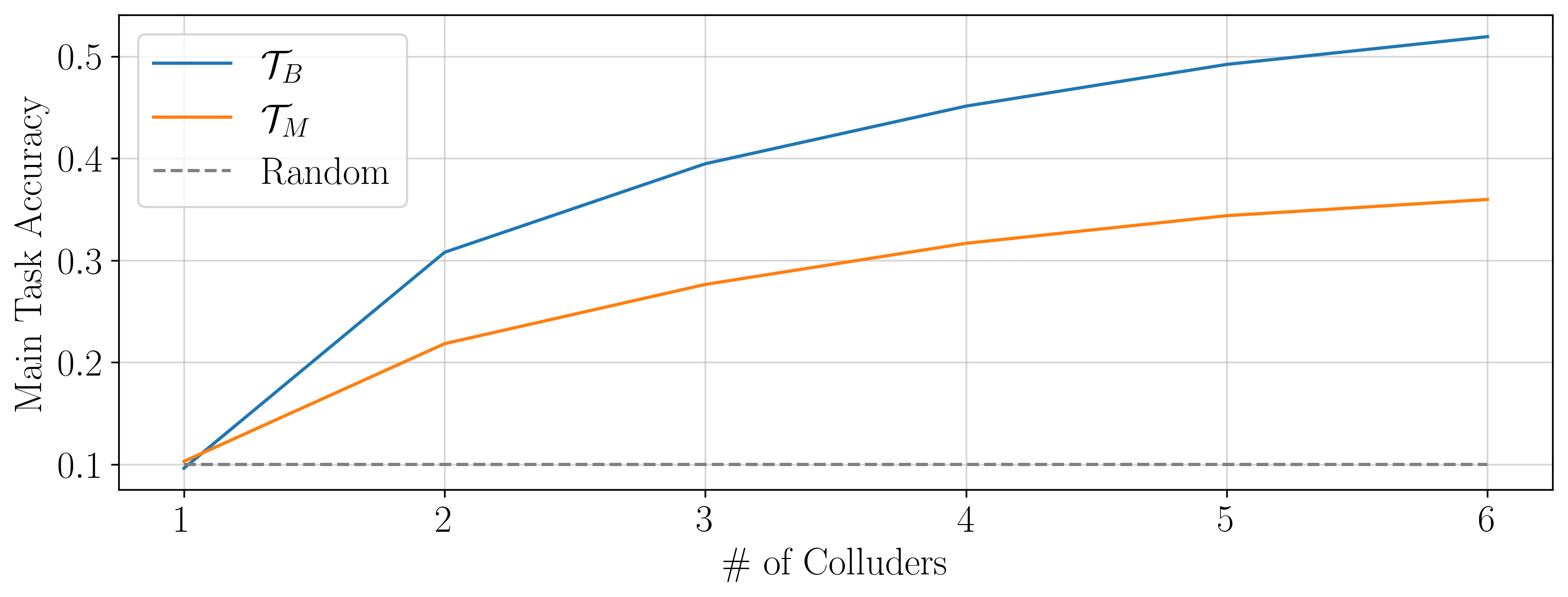}
    \caption{Evolution of the main task accuracy on $\mathcal{T}$.}
    \label{fig:triggeracc}

\vspace{\vheightsave}
\end{figure}

\begin{figure}[t]
    \centering
    \includegraphics[width=\graphwidth]{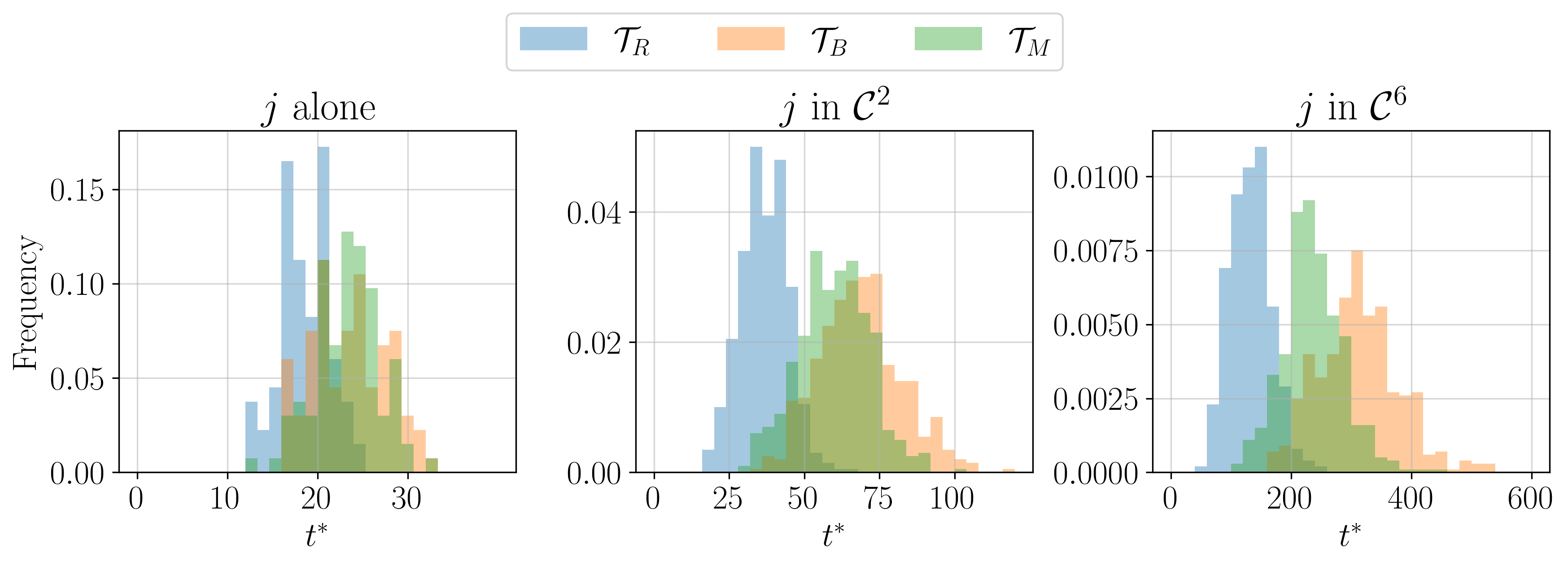}
    \caption{Experimental distribution of $t^*$ according to $\mathcal{T}$. }
    \label{fig:trigger}

\vspace{\vheightsave}
\end{figure}

Another limitation of the proposed approach, briefly discussed in Sec. \ref{sec:triggsetimp}, is that the merging of the models may skew the output towards the common main task label when triggers resemble images from the main dataset, further compromising the MA. This effect can be observed in Fig. \ref{fig:triggeracc}, and unfortunately makes benign triggers less suitable for this scheme, as one can see in the comparison of the required $t^*$, in Fig. \ref{fig:trigger}. On a more positive note, it seems that the use of $\mathcal{T}_M$ could be a possible compromise between a reasonable $t^*$ and an innocent-looking trigger.

\subsubsection{Evaluation of the False Negative Rate}

Because $\epsilon_2$ is set using $P_{\mathsf{col}}(S_j^{(i)})$, the FNR could unfortunately be higher when further attacks to the watermark are performed. However, due to the additional accusation condition from \eqref{eq:pfpbound}, no false negative was found in any of the experiments, including sampling 2500 collusions from $Z^{c_0}$ with no further attacks, finetuning, and pruning. Although this makes it difficult to estimate the FNR in a real setting, a deterrent to unlawfully leak or exploit the model still exists as long as FNR $\leq \frac{1}{2}$ \cite{Skoric12}.

\subsubsection{Simultaneous White-Box Fingerprinting}
\begin{figure}[t]
    \centering
    \includegraphics[width=\graphwidth]{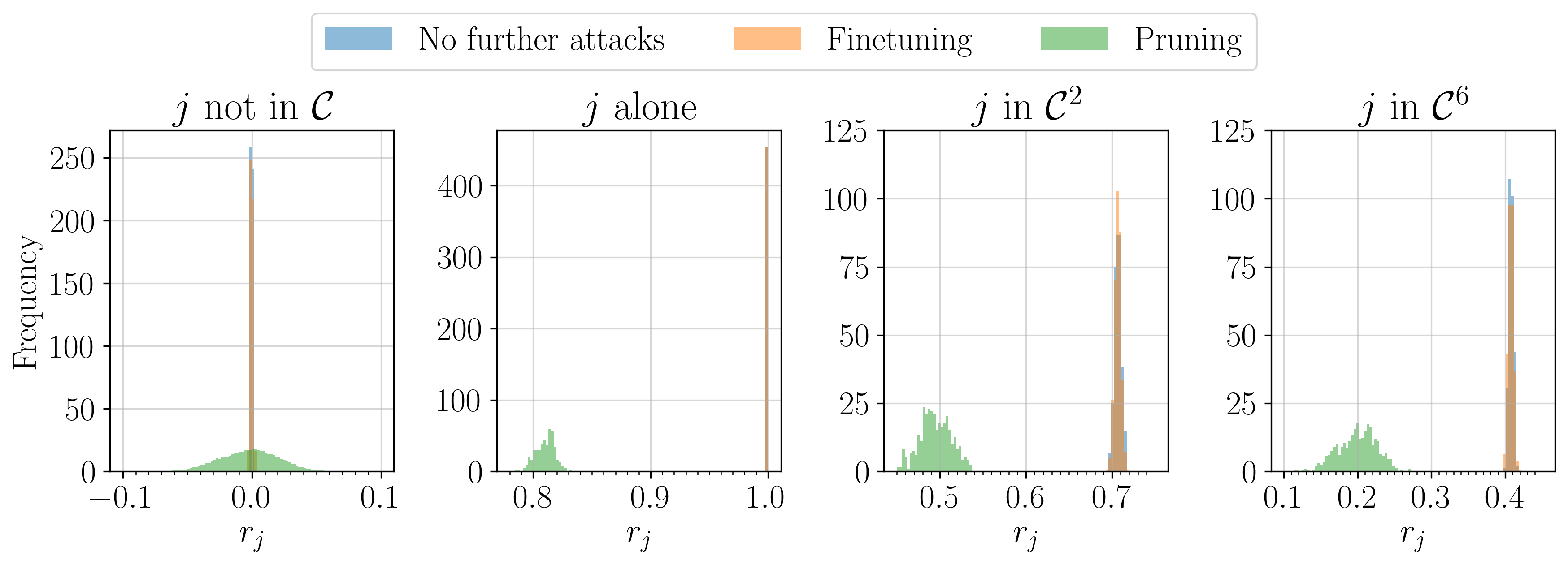}
    \caption{Experimental distribution of $r_j$.}
    \label{fig:wb}
\vspace{\vheightsave}
\end{figure}

For the white-box scheme the distribution of projections $r_j$ can be seen on Fig. \ref{fig:wb}, for different collusion sizes and further attacks. By this, statistic $r_j$ could be effectively used to accuse all guilty users (at least up to $c_0$) after the black-box queries have detected the stolen model. This accusation would be possible even after pruning, which is especially destructive, as the watermark may be exploiting weights that are not too important for the main task.\looseness=-1

\subsubsection{Effect on the Main Task}
To ensure the proposed approach was not destructive to the main task,  in Table \ref{tab:accuracymain} displays the average main task accuracy for the different schemes. Both the white and black-box schemes appear to have a small impact on the performance compared to the baseline non-watermaked model, yielding a reasonable trade-off with their traitor tracing capabilities.

\begin{table}[t] 
\caption{Influence of watermarking on the main task.}
\begin{center}
\begin{tabular}{c|c|}
\cline{2-2}
 & Accuracy on MNIST (\%) \\ \hline
\multicolumn{1}{|r|}{Non-WM Model (baseline)} &  \textbf{97.64}\\ \hline
\multicolumn{1}{|r|}{White-Box-WM Model} &  97.53 \\ \hline
\multicolumn{1}{|r|}{Black-Box-WM Model} & 97.52 \\ \hline
\multicolumn{1}{|r|}{White-and-Black-Box-WM Model} & 97.46 \\ \hline
\end{tabular}
\label{tab:accuracymain}
\end{center}
\vspace{\vheightsave}
\end{table}

\balance
\section{Discussion and Open Problems} \label{sec:discussion}

The proposed scheme opens the door to collusion resistant black-box fingerprinting, requiring significantly less queries than the existing approach, as discussed in Sec. \ref{subsec:expresult}, but it still has many limitations. 

One of the bigger challenges to be resolved is the violation of the MA, which is the backbone of the Tardos codes, as it can lead to negative scores for guilty users thus hindering the accusation process. Table \ref{tab:mav} shows the average percentage of triggers that do not hold this assumption in our experiments. Although attacks have a significant impact on the MA, even violating it 31.7\% of the time, this ends up being compensated by utilizing more triggers for the decision. This makes the accusation still possible, but it is not ideal. Further analysis on the impact that the merging of different model copies has on the triggers' output, and the bias introduced by the main task, can help understand how these violations appear, and can potentially alleviate this undesirable effect. 

\begin{table}[t]
\caption{Marking assumption violations.}
\begin{center}
\begin{tabular}{r|c|c|c|}
\cline{2-4}
  & $j$ alone & $j \in \mathcal{C}^2$ & \textbf{$j \in \mathcal{C}^6$ } \\ \hline
\multicolumn{1}{|r|}{No further attacks}  & 0.0\% & 4.3\% & 5.6\% \\ \hline
\multicolumn{1}{|r|}{Fine-tuning} & 2.3\% & 15.1\% & 15.6\% \\ \hline
\multicolumn{1}{|r|}{Pruning} & 19.0\%  & 31.7\% & 29.0\% \\ \hline
\end{tabular}
\label{tab:mav}
\end{center}
\vspace{\vheightsave}
\end{table}

Also evident is the challenge of modelling the collusion strategy over the black-box triggers. Although Tardos codes are strategy-agnostic, the scheme design can be improved to better target a certain type of attack, if known. From the results shown in Sec. \ref{subsec:expresult}, and considering the analysis in \cite{Simone11}, it seems that the merging of the models may resemble a majority-voting scenario upon the triggers, which would make sense intuitively; in any case, only model averaging collusion of small networks has been considered in this work, so further analysis would be needed to better characterize the target strategy. Also, it is worth noting that in this scenario classes might not be equiprobable, and likely depend on the type of trigger. Hence, a better understanding of the feature space, also considering other architectures and tasks, would allow for a more effective design of the codes and the score function.

\bibliographystyle{ieeetr}
\bibliography{wifs23}

\end{document}